\documentstyle[12pt]{article}
\input{psfig}
\newcommand{\be}{\begin{equation}}
\newcommand{\ee}{\end{equation}}
\newcommand{\ba}{\begin{eqnarray}}
\newcommand{\ea}{\end{eqnarray}}
\newcommand{\A}{d{\cal A}}
\newcommand{\ds}{ds^{2}}
\newcommand{\zm}{z_{max}}
\newcommand{\zt}{z_{T}}
\newcommand{\la}{\lambda}

\newcommand{\ap}{\alpha '}

\newcommand{\ra}{\rightarrow}
\newcommand{\cij}{{\cal C}_{IJ}}

\begin{document}
% title page 
\begin{titlepage}
\rightline{hep-th/9911172}
\rightline{TAUP-2604-99}
%\rightline{\today}
\vskip 1cm
\centerline{{\Large \bf  On the Supergravity Evaluation of Wilson Loop Correlators}}
\centerline{{\Large \bf in Confining Theories }}

\vskip 1cm
\centerline{ J. Sonnenschein and A. Loewy}

\vskip 1cm
\begin{center}
\em School of Physics and Astronomy
\\Beverly and Raymond Sackler Faculty of Exact Sciences
\\Tel Aviv University, Ramat Aviv, 69978, Israel
\\ cobi, loewy2@post.tau.ac.il

\end{center}
\vskip 1cm

\begin{abstract}
We explicitly show the area law behavior of a circular Wilson loop in 
confining theories from supergravity. We calculate the correlator of
two Wilson loops from supergravity in confining backgrounds. We find
that it is dominated by an exchange of a ``scalarball'' that is lighter
than the glueballs. We interpret these results in terms of the meson-meson potential in such theories.
\end{abstract}
\end{titlepage}

\section{Introduction}
In the AdS/CFT correspondence \cite{malda,witten,gubser} correlation
functions of gauge theory operators can be calculated by evaluating  the 
supergravity action on AdS$_{5} \times S^{5}$ with
the gauge theory operators positioned on the boundary of AdS$_{5}$, acting as a 
source of supergravity fields. Every gauge theory
operator has a corresponding supergravity mode. 

The correspondence also holds when probes are introduced in AdS$_{5}$. A
fundamental string with ends  on the
boundary  is one such probe. It was shown in
\cite{malda1,rey}
that the renormalized classical action of such a string  translates into
the expectation value of a Wilson loop in the dual gauge theory picture. 
Apart from  the  infinite strip loop which measures the 
potential between a quark-anti-quark pair, stringy computations were
also made for circular loops \cite{gross,malfish} and  loops with
cusps \cite{drukker}.  The  incorporation of   
quadratic quantum corrections  to the classical determination of the Wilson
loops was addressed in \cite{FGT,GrOl1, KSSW}.

 The string worldsheet acts as a source of supergravity
modes just like the gauge theory operators located on the 
boundary \cite{witten,gubser}. 
In \cite{malfish,zarembo1,callan,ulf} the supergravity action on 
AdS$_{5} \times S^{5}$ was
calculated in the presence of such a string worldsheet in order to get the
expectation value of some operator ${\cal{O}}(x)$ in the
presence of an external quark or a quark-anti-quark system. The action
was calculated by considering the exchange of the corresponding
supergravity mode from the worldsheet to the boundary point $x$. 

The Wilson loop OPE and the correlation of two loops were calculated
using the same method \cite{malfish}.
 It was shown in \cite{gross,zarembo} that there
is a solution of the string equation of motion that describes a surface
that ends on two circles. This solution is stable only when the
distance between the two loops is of the order of their radius. For longer
distances the  connecting world sheet degenerates into a thin tube
\cite{gross}.

A natural question  raised immediately after the  evaluation
of the Wilson loop  of the ${\cal N}=4$ SYM theory,  was how
to extend these results to non-supersymmetric gauge dynamics. 
Witten \cite{witten1} proposed   to describe such theories 
using  the near extremal solution, 
in the limit of   
 zero radius of the compact  Euclidean time direction  combined 
with anti-periodic boundary conditions. This scenario
 renders   the gauginos and
adjoint scalars  into very massive fields and hence resembles  the pure YM 
theory.   
This approach was utilized to determine the behavior of the 
quark anti-quark potential for
the  ${\cal N}=4$ theory at finite temperature \cite{RTY,cobi2}
 and the   ``pure YM'' theory  in three dimensions \cite{cobi1}.
Later, a similar procedure was invoked to compute Wilson loops of four
dimensional YM
theory,
 `t Hooft loops \cite{cobi1, gross},  the  potential
in MQCD and in Polyakov's type 0 model \cite{KSS1,Polyakov,kinar}.

 In \cite{kinar} a unified scheme for such models  was introduced 
 and a theorem,  that determines the leading and next to
leading behavior,   was proven and applied to several models. 
In particular a corollary
of this theorem states the sufficient conditions  for the potential to have
a confining nature. Confining Wilson loops were also discussed in 
\cite{GrOl,DorPer}.

The purpose of  this paper is to analyze the two point function of Wilson 
loops associated with  supergravity backgrounds 
that correspond to the large $N$ limit of 
 non supersymmetric confining gauge theories.
To achieve this goal we combine the approach of \cite{malfish} 
with the recipe of \cite{witten1}.
The two point function of infinite strip  Wilson loops translates
into the interaction potential between mesons composed of heavy quarks
in three dimensions. We find that the potential has the expected dependence
on $N$ and that it corresponds to an exchange of a "scalarball"
which is lighter than the glueball. A similar 
approach can also be applied to determine the potential between baryons.

The paper is organized as follows.
 In section 2 we review  the case of an
infinite strip Wilson loops  in the background of an AdS$_5$  ``black hole'',
 and calculate the
circular Wilson loop area element. We explicitly show the area law in
the circular case, and get the same string tension as in the strip
case. In section 3 we calculate the correlation of two
 Wilson loops. We  show that there are contributions from
states lighter than the glueball states of YM$_3$ \cite{oz} indicating
that the associated field theory dual is not  pure  YM  theory in three 
dimensions. 
  In section 4 we 
comment about extending the results to systems that include baryons.
The results are summarized in section 5. 

\section{Circular Wilson loops}

We start with a calculation of the area elements of circular Wilson loops in the
background of N near extremal D3 branes, also known as the AdS black
hole background \cite{witten1}.
\be ds^{2}=\ap \Big[
\frac{u^{2}}{R^{2}}(1-\frac{u_{T}^{4}}{u^{4}})dt^{2}+\frac{u^{2}}{R^{2}}dx_{123}^{2}+\frac{R^{2}}{u^{2}}(1-\frac{u_{T}^{4}}{u^{4}})^{-1}du^{2}+R^{2}d \Omega_{5}^{2} \Big] \ee
The time coordinate is  compactified on a circle of
radius $1/T$. We will consider Wilson loops in the $x_{12}$ plane, and
constant $\theta$ on the $S^{5}$. In order to make the
equations simpler we shall use $z=1/u$, and  rescale the coordinates as follows
\be \label{units} \ap R^{2}=1 \ \ \ \ \  \frac{x}{\zm R^{2}} \ra x  \ \ \
\ \ \la= \frac{\zm^{4}}{\zt^{4}} \ \ \ \ \ \frac{z}{\zm} \ra z \ee
where $\zm$ is highest point in the worldsheet, and $\zt=(\pi R^{2} T)^{-1}$. The
metric is
\be \label{metric} \ds  = \frac{1}{z^{2}} \Big[ dt^{2}(1-\la z^{4}) + dx_{123}^{2}
+ (1-\la z^{4})^{-1} dz^{2} \Big]  +   d \Omega_{5}^{2} \ee 
AdS$_{5}$ corresponds to $\la=0$. The
dilaton is independent of $\la$ and $z$. 

Infinite strip Wilson loops have been thoroughly investigated in this
background \cite{cobi1,cobi2,kinar,olsen}. The worldsheet area element
for a loop in the x-y plane is given by
\be  \label{rec} \A=\frac{1}{2 \pi \ap} \frac{dy dz}{z^{2} \sqrt{1-z^{4}} \sqrt{1-\la z^{4}}}
\ee where $dy$ is one of the spatial directions in the metric. 

Circular Wilson loops can be calculated using similar methods. In an
AdS$_{5}$ background circular Wilson loops have been investigated in
\cite{malfish,gross,drukker}. By using the conformal symmetry on the boundary of AdS$_{5}$
one can transform a Wilson line into a circle, thus getting a simple
analytic expression for the area of such a loop. 

In non conformal cases, such as in the background of N near extremal D3
branes, one must use the standard approach \cite{cobi1,cobi2}. Our
interest is circular loops in the $x-y$ plane. We shall write $dxdy$ as $rdrd\theta$, and the integrand is the same
as in the infinite strip case.
The action gives the expression for the radius of the worldsheet
as a function of $z$ \cite{witten1,cobi2}
\be r(z)= \int_{z}^{1}  \frac{z'^{2}dz'}{\sqrt{1-z'^{4}}\sqrt{1- \la
    z'^{4}}} \ee
The difference in the circular case is that the boundary condition is
$r(0)=a$, where $a$ is the radius of the loop. The area element of
the worldsheet is given by 
\be \label{cir} \A = \frac{1}{\ap} \frac{r(z)
  dz}{z^{2} \sqrt{1-z^{4}} \sqrt{1-\la z^{4}}}  \ee
When integrating this area element the divergent part is proportional
to the circumference of the loop and is subtracted to get a finite
area.  
It can be seen that upon setting $\la=0$, which is the extremal case, we get that the area is
independent of the radius as was pointed out in \cite{malfish}. In the limit
$\la \ra 1$ the dominant contribution to the area
will be from $z=1$, where we can write  \be \A=\frac{1}{\ap} r(z) d
r(z) \ee This gives the area law for circular loops 
$S=a^{2}/2$. Rescaling back the coordinates we get ($\la=1$)  \be S=\frac{1}{2}
\pi R^{2} T^{2} S_{B} \ee
where $S_{B}= \pi a^{2}$ is the area of the loop on the boundary. A
similar expression was found in \cite{cobi1} for the connection between the
energy and separation of a quark anti-quark system from the strip
loop.

For a general metric the action will be of the general form 
\be S = \frac{1}{\ap} \int r dr \sqrt{f^{2}(z)+g^{2}(z) (z')^{2}} \ee
with the boundary condition that the highest point in the worldsheet
is $z=1$.
In this case the circular loop will have the following radius, and
area
\ba \label{al} r(z)&=&\int_{z}^{1} dz' \frac{g(z')/f(z')}{\sqrt{f(z')^{2}/f^{2}(1)-1}} \nonumber \\
       \A&=&\frac{1}{\ap} \frac{r(z)g(z)f(z)}{f(1)\sqrt{f(z)^{2}/f^{2}(1)-1}} \ea 
It can be seen that if $g(z)$ is singular at $z=1$, then the dominant
contribution to the radius will come from $z=1$. Thus we can write $S$
as \be S=\frac{1}{\ap} \int f(1) r(z) dr(z)=\frac{f(1)a^2}{2\ap} \ee
in which case the area law behavior is evident. In case $g(z)$ is not singular in the $0<z<1$ interval, but $f(z)$ has
a minimum at $z=1$, the integrals in (\ref{al}) are still dominated
by the contribution at $z=1$. The leading behavior will still be an
area law for $f(1)>0$.

\section{Wilson loops correlators}

It was shown by \cite{zarembo1,gross} that when the separation of the
two loops in AdS$_{5}$ is of the order of their size, there is a solution of the
string equation of motion that describes a connected surface ending on
the two loops (fig.1a). As the separation increases the tube
shrinks, and becomes unstable. At large distances the correlation is due the exchange of
supergravity modes in the bulk between the worldsheets of the loops. 

The long distance  correlation of two circular loops was calculated in \cite{malfish} for
the ${\cal{N}}=4$ SYM theory. The long distance correlation is given
by the exchange of light supergravity modes in the bulk of AdS$_{5}$ that
couple to the worldsheet of the Wilson loops (fig.1b).
\begin{figure}[h]
\centerline{\psfig{figure=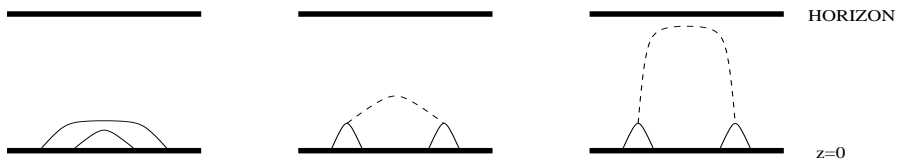,width=12cm,height=2cm,clip=}}
\caption{(a) The connected surface. (b) The exchange of supergravity
  modes between the worldsheets at distances $L \ll \zt$. (c)
  The exchange of supergravity modes at long distances.} 
\end{figure}
It seems plausible that the same picture would hold for other
backgrounds that have field theory duals. In this note we are interested in the background
created by $N$ near extremal D3 branes.

Starting with ten dimensional supergravity compactified on $W \times S^{5}$ we
may write the ten  dimensional fields as 
\be \Phi = \sum_{k,I} \phi_{k} Y_{k,I} \ee
where $\phi_{k,I}$ is a five dimensional field, and $Y_{k,I}$ are the
spherical harmonics on $S^{5}$ with total angular momentum $k$.
The spectrum of ten dimensional supergravity compactified on $S^{5}$
was investigated in \cite{kim,marcus}. In what follows we will concentrate only on the dilaton,
$\phi_{k}$, and $s_{k}$, which is a linear combination of the trace of
the metric on $S^{5}$ and the 4-form. These fields
have the following five dimensional masses
\ba \phi_{k} & m_{\phi}^{2}=k(k+4) & k \ge 0 \nonumber \\
       s_{k} & m_{s}^{2}=k(k-4) & k \ge 2 \ea
Note that the field $s_{k}$ has for $k=2,3$ a negative
$m^{2}$. However, these modes are not tachyonic, since they propagate
on a space of negative curvature. 

The general form of the correlation is  obtained by integrating on the
worldsheets the amplitude for the exchange of a supergravity mode in
the bulk between points on the worldsheets, and then summing over all modes that can be exchanged \cite{malfish}.
\be \label{cor} \log \Big[ \frac{<W(0)W(L)>}{<W(0)><W(L)>} \Big] = \sum_{i,k,I}
Y_{k,I}^{2}  \int \A_{1} \int \A_{2}
 \ f_{1}^{i,k} \  f_{2}^{i,k}  \ G^{i,k} \ee
where $f_{1}^{i,k}, \ f_{2}^{i,k}$ are the couplings of the field $i$
to the worldsheet (in general they can be functions of $z$), $k$ is
the momentum on $S^{5}$, and $G^{i,k}$ is the propagator. The sum on
$I$ follows simply from the properties of spherical harmonic on
$S^{5}$ \cite{seiberg}. 

We can read off the $N$, and $g_{s}N$ dependence of the
correlation. The propagator derived from the supergravity action will
be of order $\kappa^{2} \sim \ap^{4}g_{s}^{2}$. With our choice of
units, and the factor of $\ap^{-2}$ coming from the worldsheet area
elements, the amplitude will be of order $ \ap^{2}g_{s}^{2}=g_{s}N/N^{2}$.

To get the leading contribution we shall consider
only the lowest $k$'s possible for each field. Note that the long
distance correlation given by  (\ref{cor}) does not depend on the
relative orientation of the two loops, unlike the short distance correlation \cite{zarembo1}.
The worldsheet area elements in (\ref{cor}) are known for a general
metric \cite{kinar}. But for integrating (\ref{cor}) we need to know
the propagator and the couplings in the AdS black hole background.

The coupling of $\phi_{k}$ and $s_{k}$ in the case of a worldsheet in
AdS$_{5}$ was found in \cite{malfish}. The dilaton coupling follows
simply from the relation between the Einstein and string frame
metrics, $g_{s}=g_{E}e^{\phi/2}$. Therefore the one dilaton coupling
is $1/2$, and should be the same in the present case. The $s_{k}$
coupling was found to be $f^{s,k}=-2kz^{2}/z_{max}^2$ in AdS$_{5}$,
but the coupling in a general metric is unknown to us. The couplings of
supergravity modes in AdS$_5$ were also investigated in \cite{janik}.

The effect of  the horizon at  $z=\zt$ comes in
through the propagator, and the fact that the wave function of the
exchanged supergravity  modes should obey certain boundary conditions on the
horizon, which makes the momentum in the $z$ direction discrete, as
was shown in \cite{witten1}. At large distances, a particle exchanged
between the two worldsheets will propagate parallel to the
horizon. Therefore we assume that the full propagator will for large
distances ($L \gg \zt$) be of the form
\be G(z_1,z_2,L)=F(z_1,z_2)G_{3}(L) \ee
where $G_{3}(L)$ is the flat three dimensional propagator, and all the $z$
dependence is in the unknown function $F(z_1,z_2)$, where $z_1$ and
$z_2$ are points on the worldsheets. This agrees with
Witten's observation \cite{witten1} that the exchange of supergravity
modes in the five dimensional bulk corresponds to the exchange of field
theory modes on the three dimensional boundary. The three dimensional mass is
given by $M_{i}^{2} \sim -p_{i}^{2}/ \zt^{2}$, where $p_{i}^{2}$ are
the discrete eigenvalues of the wave operator in the $z$ direction. 

In the high temperature limit the field theory scalars and
fermions acquire masses. The dilaton which couples to
$trF^{2}$ corresponds in field theory to the glueball states $0^{++}$.
Their masses were numerically calculated in \cite{oz,nunes}. The field $s_{2}$
couples to the symmetric traceless tensor $\cij trX^{I}X^{J}$, and corresponds to a ``scalarball''. When no
external quarks, which are sources of both $\phi_0$ and $s_2$,  are
present we can restrict ourselves
correlation functions of  the massless gauge fields, and only the glueball
spectrum contributes. However, when introducing  
an external heavy quark we should consider the exchange of $s_{2}$ as well,
since the worldsheet acts as its source. This is quite clear from the
fact that the Wilson loop OPE \cite{malfish} has both $trF^2$ and
$\cij trX^I X^J$ terms. 
This spoils the identification of the $T \ra \infty$ limit with pure
QCD$_{3}$. States like $s_{2}$, which have no counterparts in QCD, should be
projected out by hand if one wishes to make such an
identification. Note that even with no heavy quarks present the
glueball masses \cite{oz} are of the same order as the scalar and
fermion masses, and hence composites of the latter do not decouple.

Although the field $s_{2}$ has $m^{2}=-4$ it can be shown numerically
that all eigenvalues are negative, meaning that the three dimensional modes
have positive $M^{2}$. The lowest mode is lighter than the lightest
glueball. This is not surprising since it originated from a five
dimensional mode with $m^{2}$ lower than the dilaton's.   
For example, the lowest eigenvalue found in \cite{oz} for the dilaton
wave function was $p^{2}=-11.59$. For  the $s_{2}$ the lowest
eigenvalue is $p^{2}=-2.3$. We denote by $M_{s}$ and $M_{\phi}$ the
lightest three dimensional masses of $s_{2}$ and $\phi_{0}$.

{\em Circular loops.\/} The correlation of two circular Wilson loops
has two phases. (i) For a
small separation distance there is a connected solution. 

(ii) For large distances the correlation is dominated by the
exchange of $s_{2}$, as was the case in AdS$_{5}$ \cite{malfish}, but
now using the 3 dimensional propagator and discrete masses we get for the lightest mass
\be  \log \Big[ \frac{<W(0)W(L)>}{<W(0)><W(L)>} \Big] \sim \frac{g_{s}N}{N^{2}} \frac{
  e^{-M_{s}L}}{L} \ee
The correlation of two circular loops in
${\cal{N}}=4$ SYM calculated using the AdS$_{5}$ bulk-to-bulk
propagator gave \cite{malfish} 
\be  \log \Big[ \frac{<W(0)W(L)>}{<W(0)><W(L)>}\Big] \sim
\frac{g_{s}N}{N^{2}} \frac{a^{4}}{L^{4}} \ee where $a$ is the radius of the loop. 

{\em Infinite strip loops in x-y plane.\/} We can also calculate the
correlation of two infinite strip loops in the x-y plane. We now have a worldsheet that is extended in one
direction. Therefore the supergravity modes can be emitted and absorbed at any
point along the $y$ direction. This means that we should write
$\sqrt{L^{2}+(y_{1}-y_{2})^{2}}$ instead of $L$. According to (\ref{cor}) we
should integrate over $y_{1}, y_{2}$. 
\be \frac{1}{Y} \log \Big[ \frac{<W(0)W(L)>}{<W(0)><W(L)>} \Big] \sim\frac{g_{s}N}{N^{2}}
K_{0}(M_{s}L) \ee where $K_{0}$ is a modified Bessel function,
which has the following asymptotic behavior \be K_{0}(M_{s}L) \approx
\sqrt{\frac{\pi}{2M_{s}L}} e^{-M_{s}L} \ee

\begin{figure}[h]
\centerline{\psfig{figure=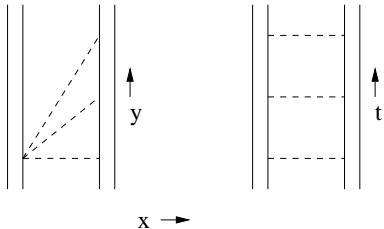,width=5cm,height=3cm,clip=}}
\caption{(a) Correlation of two strip loops in the x-y plane. (b)
  Correlation of two strip loops in the x-t plane.} 
\end{figure}

{\em Infinite strip loops in x-t plane.\/} The long distance
correlation of two infinite
strip loops in the x-t plane has the same $L$ dependence as in the
circular case. Although the worldsheet is extended in the $t$
direction, it is not a dynamical direction in the $T \ra \infty$ limit. In calculating the three
dimensional spectrum we assumed zero momentum in the
compact $t$ direction. 

The interest in the infinite strip Wilson loops comes from the fact that we
can relate the expectation value of the correlation of two such loops
to the potential between two heavy quark mesons. The potential in the
${\cal{N}}=4$ SYM case was calculated in \cite{malfish}, and was found
to be of the form expected from QCD calculation \cite{peskin}. 
\be  \label{pot} V_{mm} \sim \frac{1}{Y} \log \Big[
\frac{<W(0)W(L)>}{<W(0)><W(L)>} \Big] \ee
The infinite strip loop in the x-t plane corresponds to a meson in the
${\cal{N}}=4$ theory at finite temperature. The strip loop in the x-y
plane corresponds in the $T \ra \infty$ limit to a meson in a
``QCD$_{3}$-like'' theory. In order to make the identification with
QCD$_3$ we must project out states like $s_2$ that have no
counterparts in QCD$_3$.

For $L \gg \zt$ we get the following potentials
\ba 4d \ \ T>0 & & V_{mm} \sim \frac{g_{s}N}{N^{2}}
\frac{e^{-M_{s}L}}{L}  \\ \nonumber 
      QCD_{3} & & V_{mm} \sim \frac {g_{s}N}{N^{2}} K_{0}(M_{\phi}L) \ea 
These are the expected potentials in theories with a mass gap, and the
specified dimensionality. 

The generalization to other backgrounds \cite{cobi4} is more subtle. The
background
of other Dp-branes is not of the form $W \times S^{5}$. Therefore the
exchanged particles have Kaluza-Klein masses that vary with their
position (since the radius of the sphere is a function of $z$). One
should also make an analysis of the spectrum of type IIB supergravity
on the desired background.

\section{Baryon correlators}
The correlation of baryons can be calculated using the same
methods. 
The baryonic configuration \cite{witten2} in  the $AdS_5\times S^5$
  background is composed of a baryonic vertex
made out of a D5 brane that wraps  the $S^5$ and a set of $N$ strings stretching between the vertex and the boundary of $AdS_5$.
In  \cite{cobi3} 
a simplified picture of this system was introduced 
 where the action is just the sum  of 
the action of  the  wrapped  D5 and the $N$ strings.
 It was shown  there   
 that the  baryonic vertex will be situated at a certain 
$z_0\neq 0$ and  the  
total action was found to be  proportional to $N$ times the 
action of a corresponding Wilson loop, namely $\sim \frac{-N {\sqrt{g_sN}}}{
L}$
where $L$ is the radius of the baryon.  Improved   calculations based
on  the  BPS nature of the configuration  were performed in
\cite{BPSbaryon}.
For the baryonic configuration in the confining non supersymmetric
background
these  BPS methods  are  not applicable. A simplified 
analysis of the three dimensional 
baryons derived from the large $T$ limit of the $AdS_5$ black hole
background
 resulted in energy of the  form $\sim N(R^2T^2) L$ \cite{cobi3}.  

 Two such baryonic configurations  can exchange
supergravity modes that can be emitted and absorbed by the worldsheet
of the strings, and the  wrapped D5-brane. 
It is easy to realize that the 
 $L$ dependence of such an exchange  is  the same as in the Wilson loop
correlations. 
However, 
the $N$ dependence will now be different since each of
the baryon worldsheets is  $N$ times that of a corresponding  Wilson loop.  
The meson-baryon potential
will be of order $N^{-1}$, and the baryon-baryon potential of order
1. Again, this is the expected behavior from theories with no
dynamical quarks \cite{witten4}.

\section{Discussion}
\newcommand{\adss}{$AdS_5\times S^5$ }
The purpose of this  project was  to compute 
certain properties of confining  large $N$ gauge 
theories in the framework of  the dual 
supergravity  picture. These properties include the 
expectation value of circular Wilson loops
and the two point function of such loops and of infinite strip loops. 

Whereas  the renormalized   circular Wilson loop in the  \adss background 
was shown to be independent of the area of the 
loop \cite{malfish, gross, drukker},  the corresponding
expectation value in the  confining background admits an area law
behavior similar to
the result for  the infinite strip in that background \cite{cobi1}. 
In \cite{kinar} sufficient
conditions  for confinement where written down for a class of  generalized metrics.
In the present work we show that  indeed  loops that obey these
conditions yield an area law behavior for the circular  Wilson loop. 
It will be interesting to further generalize this result to smooth
loops of arbitrary shape and to loops that include
cusps \cite{drukker}.   

A string dual to pure three dimensional  YM theory was proposed in \cite{witten1}.
The idea is to put  the \adss background at finite temperature, impose
anti-periodic boundary conditions and  take the infinite temperature limit. 
The corresponding gauge picture is that of the ${\cal N}=4$ SYM in the infinite
temperature limit. Naively, it seems that the latter   model corresponds
to the pure YM theory  in three dimensions because the gauge fields remain massless but  the
gauginos and the adjoint scalars acquire mass  proportional to the 
temperature.  

Our present work 
 provides further evidence that in fact  this is not the 
case. The gravity background  corresponds to a confining theory but not that
of the three dimensional pure YM theory. This follows from the calculation of  the 
meson-meson interaction that is dominated by an exchange of
the  $s_2$  mode rather than the dilaton. 
The $s_2$  mode, which corresponds 
in the finite temperature ${\cal N}=4$ description to a bound state of
two scalars,
and is obviously absent from the pure YM spectrum,  is lighter 
than the dilaton that corresponds to the glueball $0^{++}$. It will be
interesting to work out the exact couplings of modes like $s_2$ to the
string worldsheet in a general background. This will enable an
explicit calculation of the correlator dependence on the size and
shape of the loop. We argue that a similar structure of exchange interaction dominates the 
potential between two baryons.

\paragraph{Acknowledgments} We would like to thank O. Aharony and
Y. Oz for their comments. This work was  supported in part by the US-Israel Binational Science
 Foundation, by GIF - the German-Israeli Foundation for Scientific Research,
and by the Israel Science Foundation.

%bibliography

\end{document}